\documentclass{article}

\usepackage{arxiv}

\usepackage{lineno,hyperref}

\usepackage{subfiles}
\usepackage{textcomp, gensymb}
\usepackage{subfig}
\usepackage{adjustbox}

\usepackage{xcolor}
\usepackage[utf8]{inputenc} 
\usepackage[T1]{fontenc}    
\usepackage{hyperref}       
\usepackage{url}            
\usepackage{booktabs}       
\usepackage{amsfonts}       
\usepackage{nicefrac}       
\usepackage{microtype}      
\usepackage{lipsum}
\usepackage{fancyhdr}       
\usepackage{graphicx}       
\usepackage{natbib}
\usepackage{doi}

\hypersetup{
pdftitle={Comparative analysis of deep learning approaches for AgNOR-stained cytology samples interpretation},
pdfsubject={arxiv},
pdfauthor={João G. A. Amorim},
pdfkeywords={Deep Learning; Computer vision; Cytology; AgNOR; Cervical cancer;},
}

\title{Comparative analysis of deep learning approaches for AgNOR-stained cytology samples interpretation}

\author{ \href{https://orcid.org/0000-0003-3361-6891}{\includegraphics[scale=0.06]{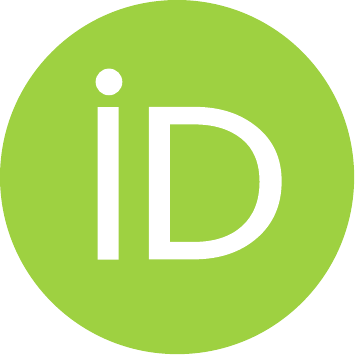}\hspace{1mm}Jo{\~{a}}o Gustavo Atkinson Amorim}\\
	Department of Informatics and Statistics\\
	Universidade Federal de Santa Catarina\\
	  Florianópolis, SC, Brazil \\
	\texttt{joao.atkinson@posgrad.ufsc.br} \\
	\And
	\href{https://orcid.org/0000-0003-0268-0233}{\includegraphics[scale=0.06]{orcid.pdf}\hspace{1mm}Andr{\'{e}} Vict{\'{o}}ria Matias} \\
	Department of Informatics and Statistics\\
	Universidade Federal de Santa Catarina\\
	  Florianópolis, SC, Brazil \\
	\texttt{andre.v.matias@posgrad.ufsc.br} \\
	\And
	\href{https://orcid.org/0000-0001-8979-4983}{\includegraphics[scale=0.06]{orcid.pdf}\hspace{1mm}Allan Cerentini} \\
	Department of Informatics and Statistics\\
	Universidade Federal de Santa Catarina\\
	  Florianópolis, SC, Brazil \\
	\texttt{allan.c@posgrad.ufsc.br} \\
	\And
	\href{https://orcid.org/0000-0001-6127-4863}{\includegraphics[scale=0.06]{orcid.pdf}\hspace{1mm}Luiz Antonio Buschetto Macarini} \\
	Department of Automation and Systems\\
	Universidade Federal de Santa Catarina\\
	  Florianópolis, SC, Brazil \\
	\texttt{luiz.buschetto@posgrad.ufsc.br} \\
	\And
	\href{https://orcid.org/0000-0002-3833-8694}{\includegraphics[scale=0.06]{orcid.pdf}\hspace{1mm}Alexandre Sherlley Onofre} \\
	Clinical Analyses Department\\
	Universidade Federal de Santa Catarina\\
	  Florianópolis, SC, Brazil \\
	\texttt{alexandre.onofre@ufsc.br} \\
	\And
	\href{https://orcid.org/0000-0003-4857-5770}{\includegraphics[scale=0.06]{orcid.pdf}\hspace{1mm}Fabiana Botelho Onofre} \\
	Clinical Analyses Department\\
	Universidade Federal de Santa Catarina\\
	  Florianópolis, SC, Brazil \\
	\texttt{fabiana.onofre@ufsc.br} \\
	\And
	\href{https://orcid.org/0000-0003-4532-1417}{\includegraphics[scale=0.06]{orcid.pdf}\hspace{1mm}Aldo von Wangenheim} \\
	Clinical Analyses Department\\
	Brazilian Institute for Digital Convergence\\
	  Florianópolis, SC, Brazil \\
	\texttt{aldo.vw@ufsc.br} \\
}

\renewcommand{\shorttitle}{Comparative of deep learning approaches for AgNOR-stained cytology images}

\begin{document}
\maketitle

\begin{abstract}
    Cervical cancer is a public health problem, where the treatment has a better chance of success if detected early. The analysis is a manual process which is subject a human error, so this paper provides a way to analyze argyrophilic nucleolar organizer regions (AgNOR) stained slide using deep learning approaches. Also, this paper compares models for instance and semantic detection approaches. Our results show that the semantic segmentation using U-Net with ResNet-18 or ResNet-34 as the backbone have similar results, and the best model shows an IoU for nucleus, cluster, and satellites of 0.83, 0.92, and 0.99 respectively. For instance segmentation the Mask R-CNN using ResNet-50 performs better in the visual inspection and has a 0.61 of the IoU metric. We conclude that the instance segmentation and semantic segmentation models can be used in combination to make a cascade model able to select a nucleus and subsequently segment the nucleus and its respective nucleolar organizer regions (NORs).

\end{abstract}



\section{Introduction}
    
    Cervical cancer is a public health problem, being the fourth most common cancer type in women. The treatment has a better chance of success if cervical cancer is detected early. So the high mortality rate, fourth in women, is directly related to the non-existent or inadequate screening which leads to late detection. The main aggravation of this lesion is the Human Papillomavirus (HPV), which mainly affects underdeveloped countries \cite{gstoday2020}.
    
    Through the analysis of the cells (cytology), it is possible to diagnose cervical cancer. For this diagnostic, it is necessary to use a marker in the screening to describe the cells. The increased protein synthesis observed in aneuploid cells is one factor that characterizes the transformation of cells from normal to malignant. This way, Argyrophilic Nucleolar Organizer Regions (AgNORs) can be used as a marker for the quantification of cell proliferation, differentiation, and malignant transformation \cite{ruschoff1989nucleolar}. The AgNOR staining technique shows the potential for diagnostic cervical cancer. One of the diagnosing methods is the counting of Nucleolar Organizer Regions (NORs) \cite{Sakai2001, GarcaVielma2015}.
    
    NORs are loops of DNA in the nucleus of the cells that have ribosomal RNA genes present. Further, the nuclear proteins (that are argyrophilic) envelope the NORs, which the AgNOR staining technique highlights. In some studies, like in \cite{Onofre2008, pedrini2020analise}, the NORs are classified in silver‐stained dot aggregations or partly disaggregated nucleoli (clusters) treated as one structure, and in individual dots (satellites) outside the clusters of silver‐stained structures. This diagnostic/classification in slides stained with the AgNOR technique is done manually, which makes the process prone to human errors. The cause of these human errors is the different levels of knowledge and different visual perceptions between cytologists \cite{Li2017}. An automatic analysis (or description) of the AgNOR images can reduce these errors and speed up the process. 
    
    Even though the automated quantitative analysis of cytological images employing artificial intelligence methods has been explored for more than 25 years now, there is still no simple, automated method defined \cite{Kolles93, Kolles94}. As shown in \cite{Meijering2012, rslcyto} deep learning methods were used in most works in the computer vision field in the last years. In particular, the Convolutional Neural Networks (CNNs) have succeeded in solving problems that resisted many attempts of artificial intelligence researchers \cite{LeCun2015}.

    Although no work in the last years has been found using computer vision for AgNOR image analysis, deep learning techniques have been commonly used for the study of cytology images. The most widely used network architecture for solving segmentation problems is based on U-Net \cite{Ronneberger_2015}. The approach used in \cite{Deng2020} was divided into two steps: in the first, Pap Smear images were segmented into nuclei, cytoplasm, and background, employing a U-net model using a ResNet as its backbone and in the second step, the nuclei were classified into sub-types. Another approach technique is to use instance segmentation like in \cite{Sompawong2019}, where the authors used a Mask R-CNN model for nuclei detection in Pap smear slides.

\subsection{Objectives}
    
    
    
    In a previous study \cite{AtkinsonAmorim2020}, we published a comparative study of semantic segmentation for multiple hyperparameters to train a U-Net model with ResNet-18 as a backbone. We also have made a public dataset available with 2,540 images of fields stained with the AgNOR technique. As there are no other studies with computer vision approaches to support quantitative cytology and diagnosis of cancer on AgNOR images, this previous study will be used as a benchmark. The best result shows a 0.68 for the mean of Intersection over Union (mIoU) and 0.87 for restrictive mIoU (without background pixels).
    
    Therefore, this paper presents a comparison of approaches using deep learning for AgNOR described in cytology exams aiming the automatic image analysis. This work performed two approaches (semantic segmentation and instance segmentation) on fields generated from whole slide images. Our results and conclusion may provide some insights into what approach is better to early detect cancer using AgNORs.
    
\section{Related Works}
    A Systematic Literature Review (SLR) was executed in early 2020 with the objective of finding what are the current computer-assisted or artificial intelligence-based approaches in computer vision for the support of quantitative cytology and diagnosis of cancer in cytological exams \cite{rslcyto}. The studies were analyzed from the beginning of 2016 until  mid-2020. Based on the analysis of the papers, deep learning approaches (mostly CNN approaches) were the most common in these works. However, classical approaches such as Support Vector Machines, Random Forests, Artificial Neural Networks, super-pixel segmentation, and others, are still employed in some of the works. In this SLR it is shown that there are more studies using deep learning approaches for computer vision in recent years. When compared with the results of an SLR from a few years ago \cite{Meijering2012}, this becomes more evident. In \cite{rslcyto} the Pap Smear was the most recurrent stain in the experiments and no papers present experiments with the AgNOR technique.
    
\section{Methodology}

\subsection{Dataset}
    All data used comes from fields of slides from three different patients presenting cervical cancer, and all data came from the \textit{CCAgT dataset: Images of Cervical Cells with AgNOR Stain Technique}
    \cite{AtkinsonAmorim2020, CCAgT2020}. A total of 2,540 images with at least one nucleus labeled were used. Beyond the nucleus, the dataset has satellites and clusters labeled. The exams were realized in University Hospital Professor Polydoro Ernani de São Thiago of the Federal University of Santa Catarina (HU-UFSC)
    and the images were captured by a ZEISS Axio Scanner.Z1 with a Hitachi HV-F202SCL as an imaging device.
    
    Each image in the dataset has a resolution of 1600x1200 pixels, where each pixel has a size of $0.111 \mu m \times 0.111\mu m$. A total of 4,515 nuclei, 2,171 satellites, and 10,025 clusters are present in the dataset and they have an average area of 4,857, 100, and 125 respectively in pixels. The satellites are labeled as single points and, because of this, the area does not present a standard deviation. The nucleus and clusters have 3,109 and 80 in pixels for the standard deviation of their area respectively. These images were randomly split into training (70\%), validation (15\%), and test (15\%) sets to train and evaluate the deep learning models. 
    
\subsection{Models}
In this work, we employed two different approaches: Semantic Segmentation and Instance Segmentation. We trained the networks using NVIDIA P100 (16 GB VRAM) graphics cards, with an Intel(R) Xeon(R) CPU @ 2.20GHz $\times$ 4, 26 GiB of RAM and Ubuntu 18.04.3 LTS 64-bit using the Google Colab Platform \cite{Bisong2019}.

\subsubsection{Semantic Segmentation}
    Semantic segmentation was performed employing the \textit{fast.ai} \cite{Howard2020} framework. We used the UNet model with ResNet-18 and ResNet-34 as the backbone in two different experiments. All experiments applied the same data augmentation pipeline on the images. We applied a random flip with 50\% of probability, a random rotation with 75\% of probability (rotation between -10 to 10 degrees), zoom up at $1.1\times$, lightning, and contrast up to 20\% and symmetric warp of magnitude between $-0.20$ and $0.20$. All the training data was normalized by the ImageNet stats \cite{Deng2009}. Neither data augmentation or normalization was applied to the test set as we used it to evaluate the generalization capability of the models. We used all the classes available in the dataset in all semantic segmentation experiments.
    

    Initially, we performed an experiment using ResNet-18 as the backbone without pre-training, where we employed the images in the original tiles (1600x1200). The training process occurs in a multi-resolution process. This multi-resolution strategy allows the model to converge along with the levels (re-samples of sizes) for the best results. This strategy acts as a pre-training of the model, helping the network to learn to extract the most important features. The outline of this strategy was presented by Jeremy  Howard as an informal communication during a CNN lecture available at \cite{howard2019l3}, and in this experiment, we trained the models at the resolution of one-quarter of the original (400x300), half of the original (800x600), and on a full size (1600x1200). Each level of resolution was first trained for ten epochs in the encoder part, and then for five more epochs using the whole UNet model.


    Other experiments were performed using smaller tiles. For this, the original tiles of 1600x1200 were split into sixteen tiles of 400x300. We performed these experiments because in several of the original tiles there is only one nucleus labeled, which indicates a region of interest of less than 0.3\% in these cases. Splitting the images, it is possible to remove several smaller tiles that do not have at least one nucleus. This makes the training process faster and possibly makes the models more specialized. We used a total of 4,818 images with at least one nucleus with a resolution of 400x300 pixels. In this dataset, we performed the experiments using ResNet-18 and ResNet-34 as the backbone without pre-training, using a multi-resolution process. The training process uses half of the tile size (400x300), and on a full size (800x600) where each level of resolution was trained equals the model with the original tile size.

\subsubsection{Instance Segmentation}
    For instance segmentation, we employed the Mask R-CNN model with ResNet-50 or ResNet-101 as the backbone. The model was implemented in \textit{detectron2} \cite{wu2019detectron2} framework. No data augmentation was employed.
    
    Both experiments were trained and tested with the same parameters, using only the original tile size (1600x1200). We used ResNet and FPN backbones with standard convolutions and fully-connected heads for mask and box prediction, respectively. This combination obtains the best speed/accuracy trade-off in the experiments with COCO dataset \cite{lin2015microsoft}. To compare different models, we evaluated the results using both ResNet-50 and ResNet-101 pre-trained in the COCO dataset. They were trained by a thousand interactions using a learning rate of 0.005. Also, to help in the boundary box suppression, in the validation data we performed a selection of thresholds to scores and Intersection over Union (IoU) values. The score values are the confidence of the model in the results (from 0 to 1), and the IoU threshold is used to determine if the results are True Positives, as explained in the next section (Evaluation methods and metrics).
    
\subsection{Evaluation methods and metrics}

To evaluate the model's performance we used three approaches in a distinct fold (test fold) of the data. For the first approach, we used the accuracy metric to analyze the quality of hits in object detection. A second analysis in pixel level was used to indicate the overall quality of the segmentation approach. At last, visual and manual analysis was performed to validate the best model. 

First, it is necessary to explain the parameters (TP, TN, FP, and FN) utilized in the accuracy calculus. These parameters are usually used in the object detection approach and were used in this study to check the quality of hits at the objects/entities level:
    
\begin{itemize}
    \item \textbf{True Positive (TP)}: It is considered a TP when the IoU between the predicted object and the ground truth is higher than a given threshold (usually $IoU>0.5$)

    \item \textbf{True Negative (TN)}: TN is not used in object detection because it usually makes no sense to train a model with a negative case, or ``background", object.
    
    \item \textbf{False Positive (FP)}: When the model detects an object that is not present in the ground truth or it is present but the IoU is below a given threshold (usually 0.5).
    
    \item \textbf{False Negative (FN)}: When the object is present in the ground truth and the model does not detect it.
\end{itemize}
    
\textbf{Accuracy}: is the ratio between the \textit{True Negatives} and the sum of \textit{False Negatives} and \textit{True Negatives} \cite{Fawcett2006}, as shown in the Equation \ref{eq:acc}. It indicates the overall effectiveness of the model, and in this work utilized to analyze the quality of hits at the objects/entities level.
\begin{equation}\label{eq:acc}
    Accuracy = \frac{TP + TN }{FP + FN + TP + TN}
\end{equation}

\textbf{Intersection Over Union (IoU)}: this is a metric for semantic segmentation and a measure based on the Jaccard Index \cite{Jaccard} that evaluates the overlap between two regions in the image: the \emph{ground truth region} and the \emph{predicted region} \cite{ulku2019survey}. For pixel-wise approaches, this measure is the intersection of the pixel-wise classification results with the ground truth, to their union and this is the way used to measure the accuracy in a model of detection objects or semantic segmentation. As shown in Equation \ref{eq:iou}, $n_{jj}$ is the pixels both classified and labeled as class $j$. Similarly, $n_{ff}$ is the pixels both classified and labeled as class $f$, the correct rejection. For false cases, $n_{ij}$ is the pixels that are labeled as $i$ class but classified as $j$. Finally, $n_{ji}$ is the pixels that are labeled as $j$ class but classified as $i$
 \cite{Fawcett2006}. 
\begin{equation}\label{eq:iou}
    IoU = \sum_{j=1}^{k}{\frac{n_{jj}}{n_{ij}+n_{ji}+n_{jj}}}, \quad i\neq j
\end{equation}

\textbf{Visual inspection:} this process was made to check if the best model indicated by the metrics also is the best model by the visual analysis. This process follows the same standard, making it possible to compare the same samples for different models. Thus, it is possible to identify the most recurrent faults for each case.

\section{Results and discussion}

    All experiments were performed on the test fold, and are summarized in Table \ref{tab:results}. Despite the lower accuracy values, semantic segmentation models present better IoU results. This is probably because there is no morphological post-processing, and some objects that are not fully joined may be counted, making the metric results biased. Something similar occurs with the Mask R-CNN models where the ResNet-101 has a lower accuracy but presents a better IoU than ResNet-50. This occurs because ResNet-101 presents more than one detection for the same object so has larger numbers of FP.
    
    \begin{table*}[htp]
    \caption{Performance of each model}
    \label{tab:results}
    \begin{adjustbox}{width=\textwidth}
        \begin{tabular}{|cll|c|c|c|c|c|c|}
            \hline
            \textbf{}                                                                   & \multicolumn{1}{c}{\textbf{}}          &                                                                                & \multicolumn{3}{c|}{\textbf{Accuracy \%}}                & \multicolumn{3}{c|}{\textbf{IoU \%}}                     \\ \hline
            \multicolumn{1}{|c|}{\textbf{Model}}                                        & \multicolumn{1}{c|}{\textbf{Backbone}} & \multicolumn{1}{c|}{\textbf{Obs.}}                                             & \textbf{Nucleus} & \textbf{Cluster} & \textbf{Satellite} & \textbf{Nucleus} & \textbf{Cluster} & \textbf{Satellite} \\ \hline
            \multicolumn{1}{|c|}{UNet}                                                  & \multicolumn{1}{l|}{ResNet-18}         & Full size                                                                      & 43.3             & 52.2             & 35.0               & 83.4             & 89.3             & 94.8               \\ \hline
            \multicolumn{1}{|c|}{UNet}                                                  & \multicolumn{1}{l|}{ResNet-18}         & 1/4 size                                                                       & 54.0             & 54.5             & 34.5               & 82.6             & 91.4             & 97.4               \\ \hline
            \multicolumn{1}{|c|}{UNet}                                                  & \multicolumn{1}{l|}{ResNet-34}         & 1/4 size                                                                       & 48.8             & 56.5             & 27.6               & 83.5             & 92.0             & 98.8               \\ \hline
            \multicolumn{1}{|c|}{\begin{tabular}[c]{@{}c@{}}Mask \\ R-CNN\end{tabular}} & \multicolumn{1}{l|}{ResNet-50}         & \begin{tabular}[c]{@{}l@{}}Full size and \\ 75\% of \\ score min.\end{tabular} & 60.2             & -                & -                  & 61.94 & -                & -                  \\ \hline
            \multicolumn{1}{|c|}{\begin{tabular}[c]{@{}c@{}}Mask\\  R-CNN\end{tabular}} & \multicolumn{1}{l|}{ResNet-101}        & \begin{tabular}[c]{@{}l@{}}Full size and \\ 90\% of\\ score min.\end{tabular}  & 52.3             & -                & -                  & 70.26 & -                & -                  \\ \hline
        \end{tabular}
    \end{adjustbox}
\end{table*}

\begin{figure*}[!htp]
        \centering
        \subfloat[Original]{\includegraphics[width=0.65\textwidth]{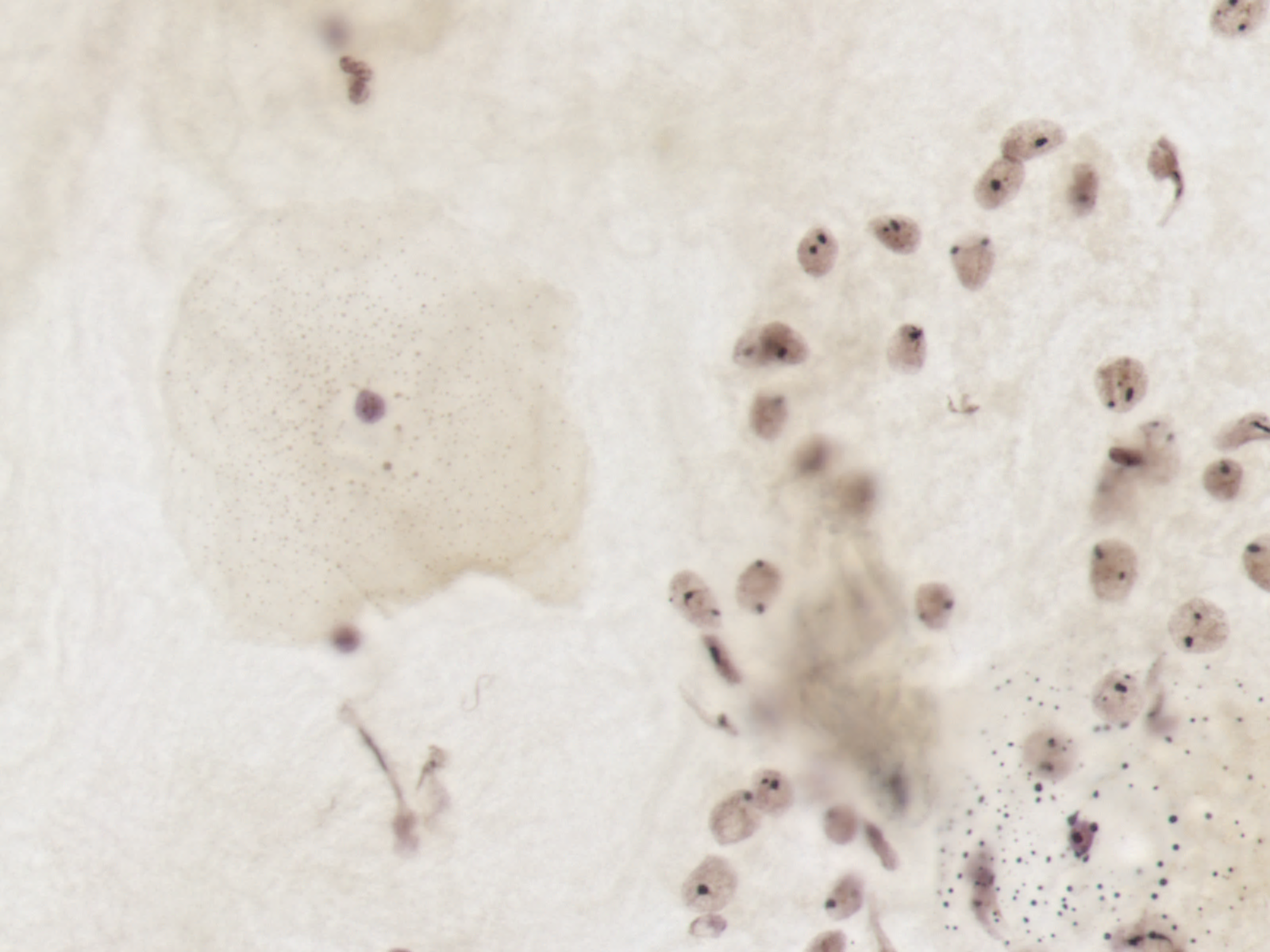}%
        \label{fig:SSORI}}
        \hfil
        \subfloat[Ground truth]{\includegraphics[width=0.4\textwidth]{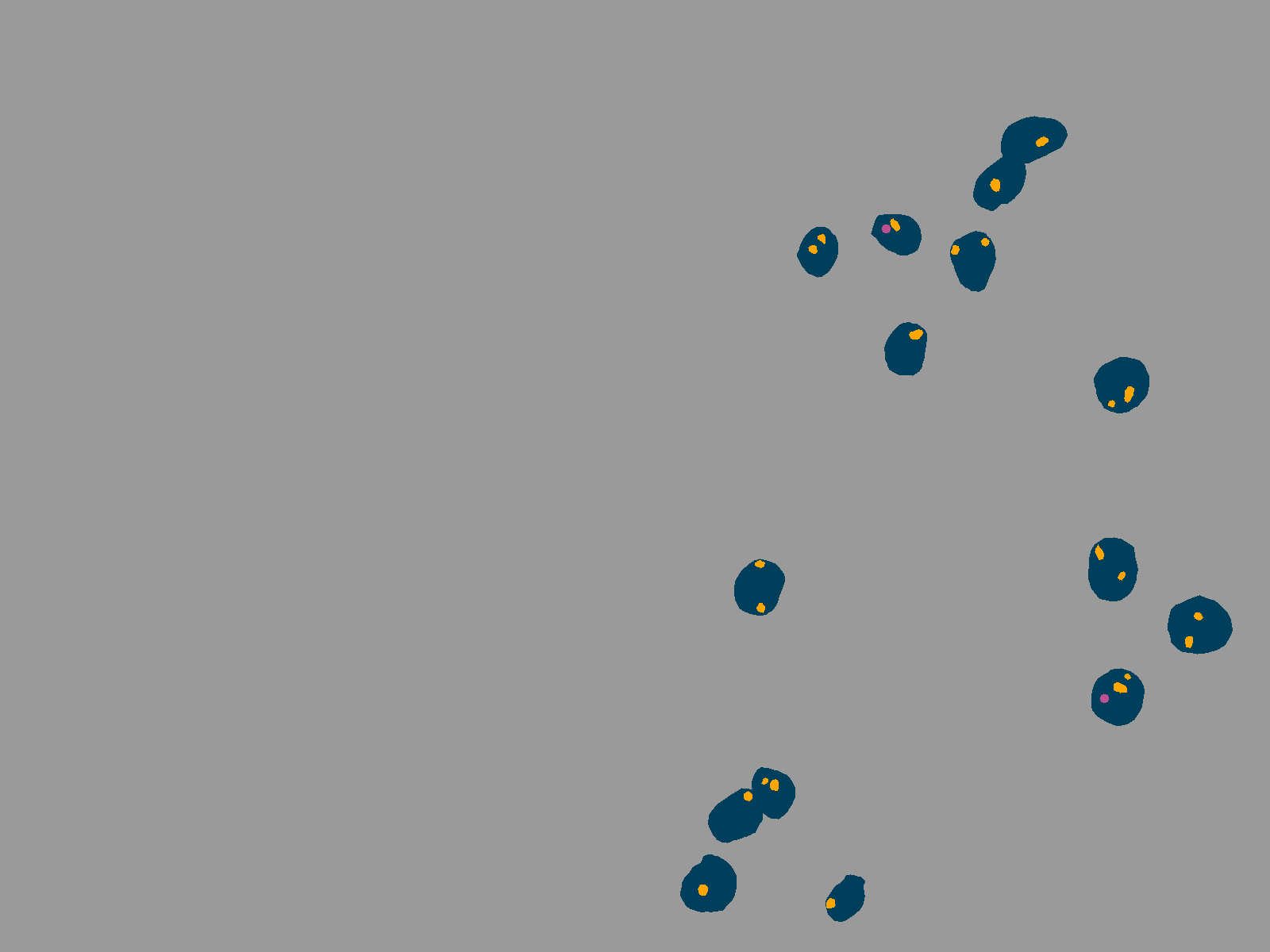}%
        \label{fig:SSGT}}
        \hfil
        \subfloat[ResNet-18 original size]{\includegraphics[width=0.4\textwidth]{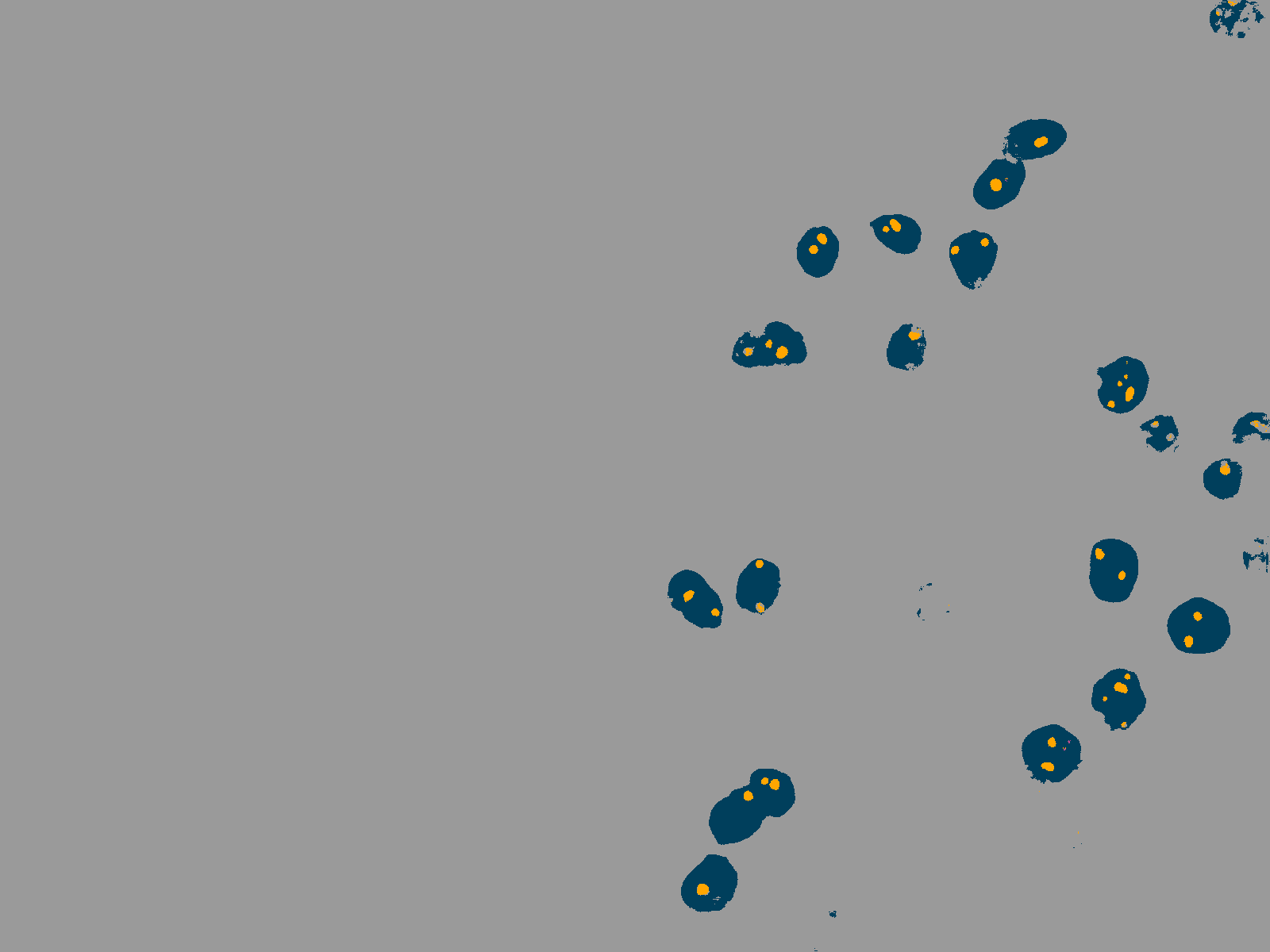}%
        \label{fig:SSr18ori}}
        \hfil
        \subfloat[ResNet-18 with tiles]{\includegraphics[width=0.4\textwidth]{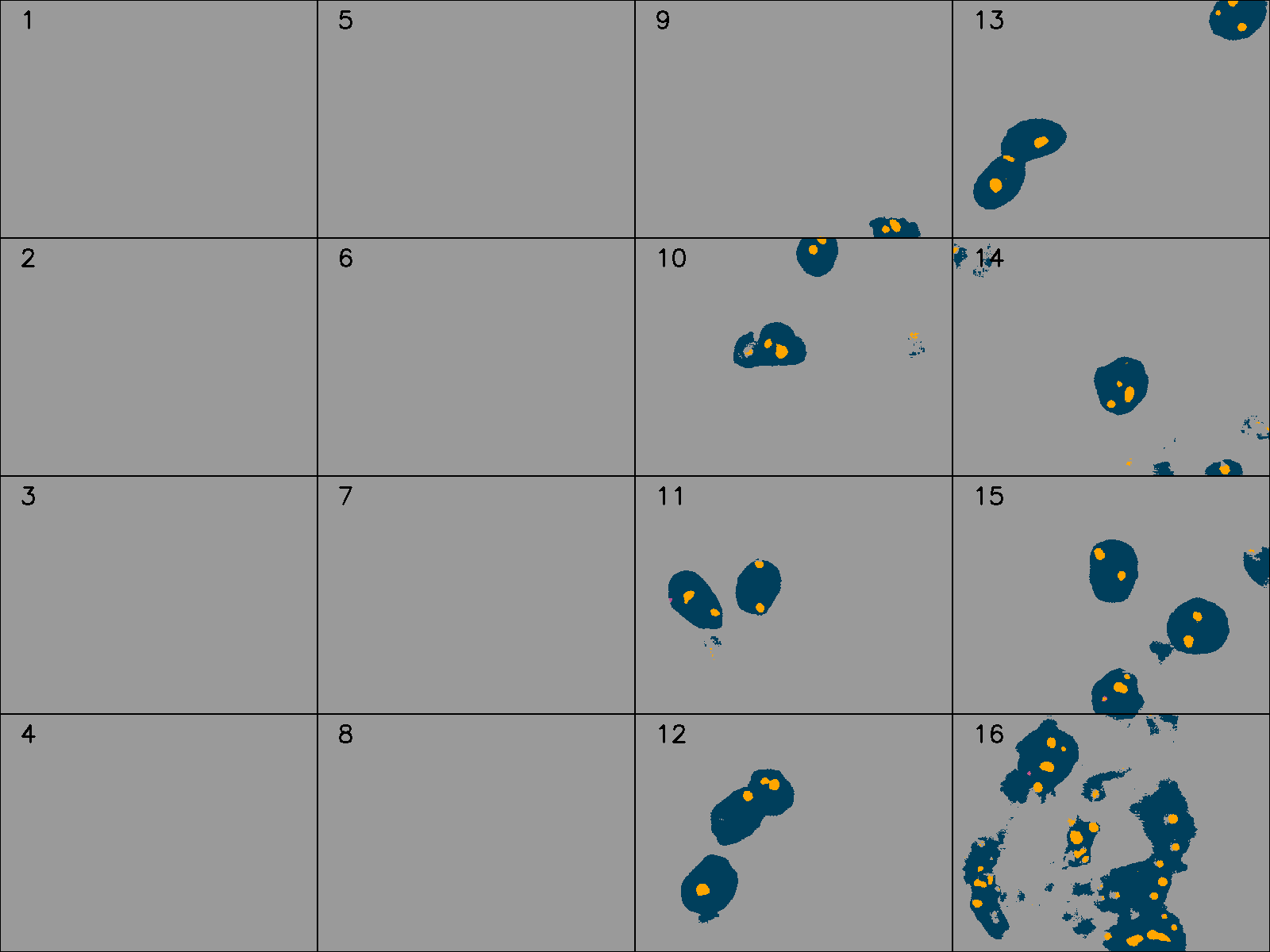}%
        \label{fig:SSr18tiles}}\hfil
        \subfloat[ResNet-34 with tiles]{\includegraphics[width=0.4\textwidth]{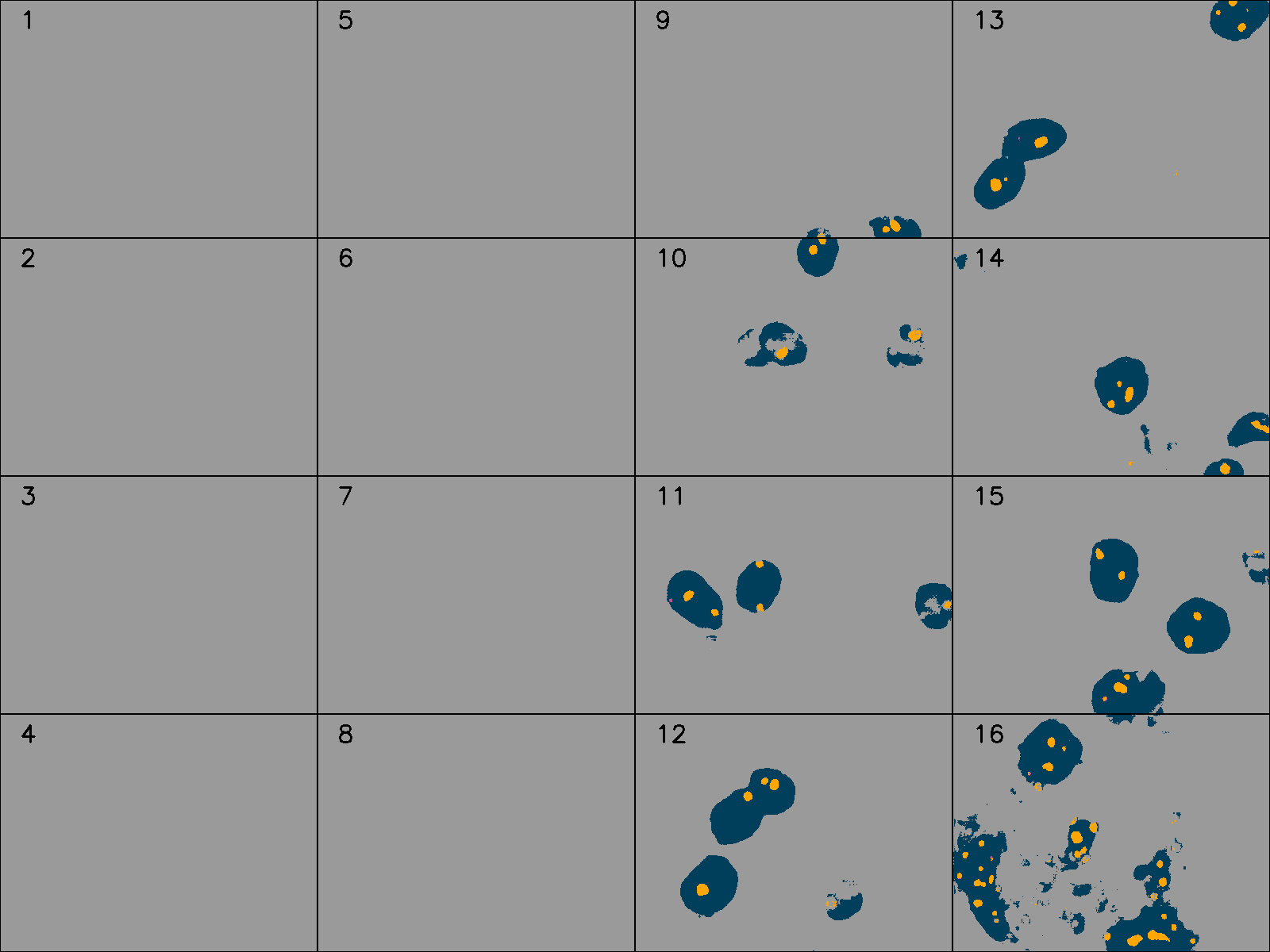}%
        \label{fig:SSr34tiles}}
        
        \caption{Results in semantic segmentation with different backbones}
        \label{fig:SS}
    \end{figure*}

    Comparing the semantic segmentation results shown in Table \ref{tab:results} and Figure \ref{fig:SS}, the models using the tiled images (400x300 size) present betters values of accuracy and IoU compared to the model using the original image size. Although the model with ResNet-34 and using tiled images present lower values of accuracy, this model shows better results with a mean IoU of 91.43. When this technique is evaluated by the visual inspection, as shown in Figures \ref{fig:SSr18ori}, \ref{fig:SSr18tiles}, and \ref{fig:SSr34tiles}, the models using tilled images detects artifacts/noise in the image, where the model using the original size ignores these noises.
    
    These experiments and comparisons show the necessity of using tiles with overlapping edges to ensure that no items of interest are separated and therefore not segmented correctly. It also shows a requirement for more elaborate post-processing to ensure correct evaluation of the slide in the case of models that use tiled images, despite they being simpler and faster. Therefore, the model using the size of the original tile (1600x1200) shows a better performance to ignore the noises and out-of-focus cases in the images. In general, all models have the same problem reported in \cite{AtkinsonAmorim2020}: they can't segment the satellites correctly, besides segmenting NORs inside the nucleus well.
    
\begin{figure*}[!htp]
        \centering
        \subfloat[Ground Truth]{\includegraphics[width=0.65\textwidth]{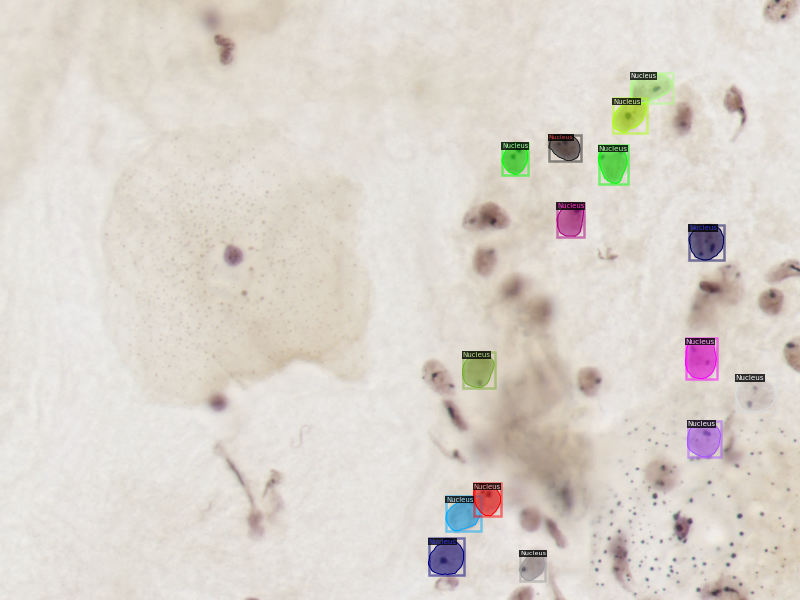}%
        \label{fig:ISgt}}
        \hfil
        \subfloat[ResNet-50]{\includegraphics[width=0.4\textwidth]{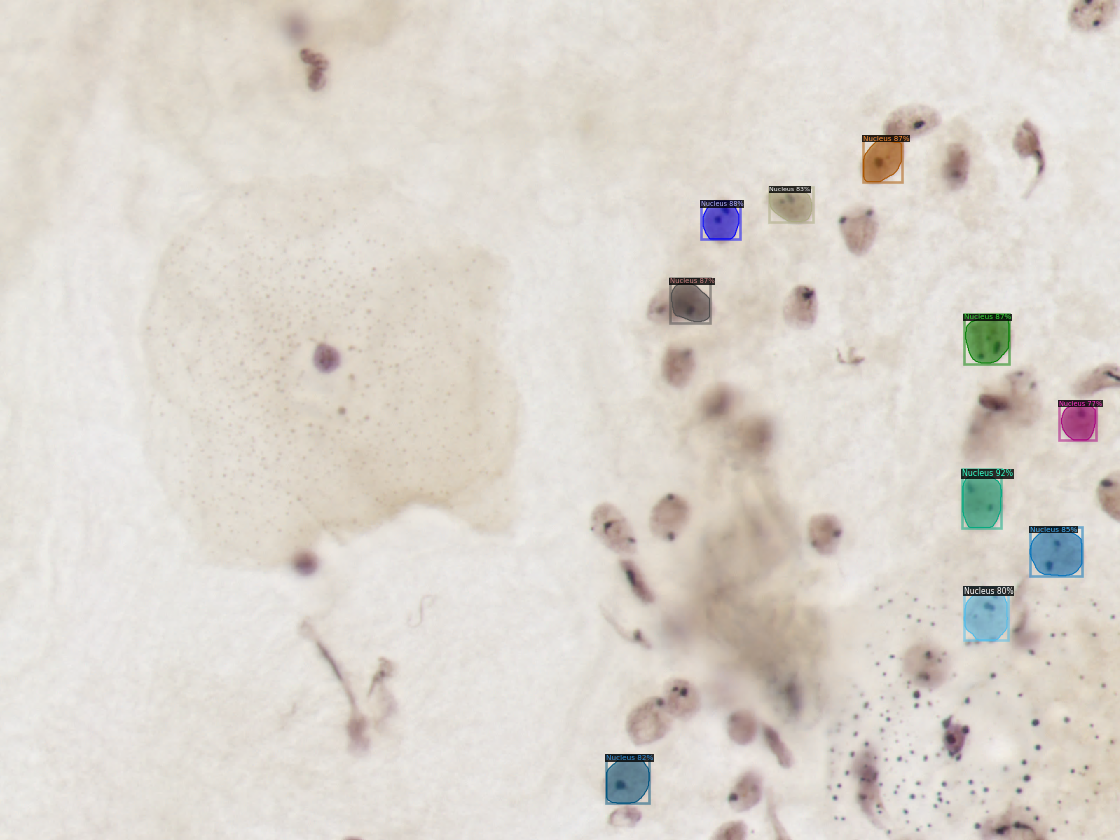}%
        \label{fig:ISr50}}
        \hfil
        \subfloat[ResNet-101]{\includegraphics[width=0.4\textwidth]{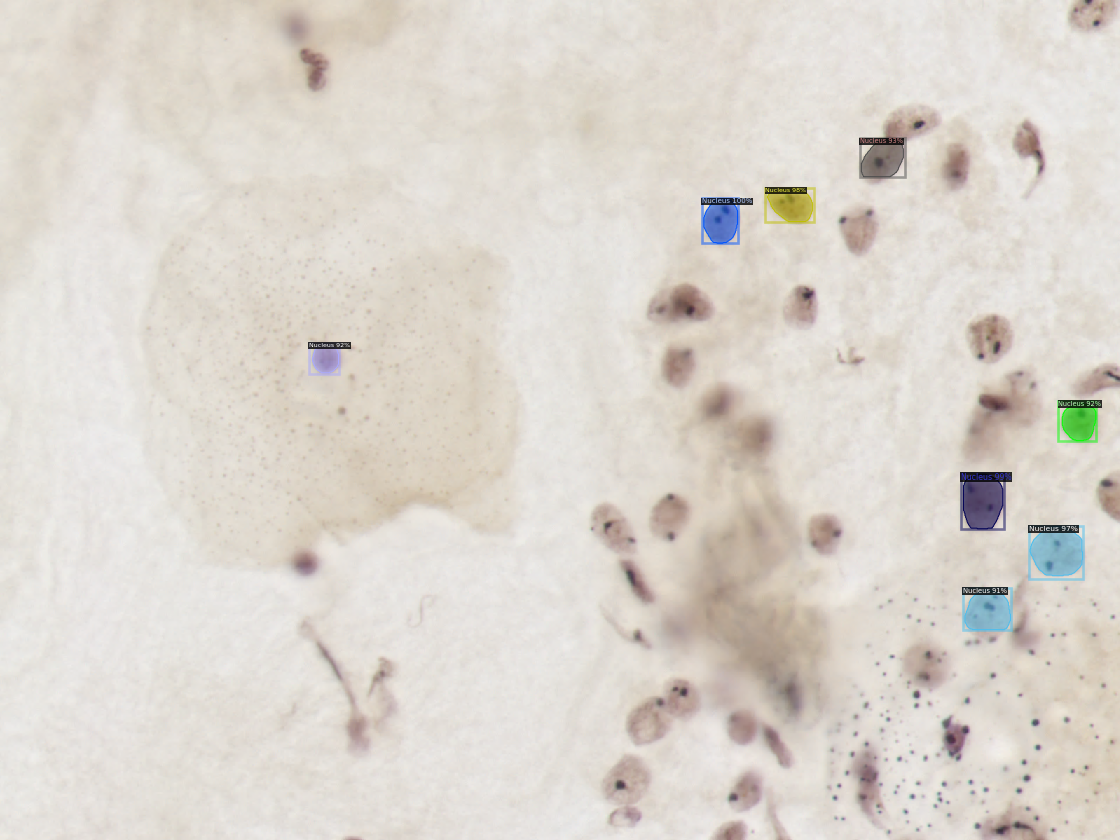}%
        \label{fig:ISr101}}
        \caption{Results in instance segmentation with different backbones}
        \label{fig:IS}
    \end{figure*}

    The results of instance segmentation are shown in Table \ref{tab:results} and Figure \ref{fig:IS}. The model with the ResNet-50 presents better results of accuracy, and the model with the ResNet-101 presents better results in terms of IoU metrics. Due to the complex background, there are still problems with small object detection. Because of that, we performed the detection of the nucleus using only the Mask R-CNN model. Because of this, only the detection of one class (nucleus) was performed with success. When this technique is evaluated by the visual inspection, as shown in Figures \ref{fig:ISr50}, and \ref{fig:ISr101}, the model with the ResNet-50 as backbone performs better than the model with the ResNet-101 as the backbone.

    The example tile, shown in the Figures \ref{fig:SS} and \ref{fig:IS} presents a high amount of nuclei, nuclei out of focus, some unusable nuclei, and also parts with artifacts and noise. In the ground truth of the CCAgT dataset shown in Figure \ref{fig:SSGT} and \ref{fig:ISgt}, just nucleus with NORs are labeled.  Like in this example, other tiles in CCAgT dataset show this same problem, where there are some items with these characteristics but not labeled.


\subsection{What is new here in relation to our previous works?}

    In comparison with our previous works, in this paper, we applied instance segmentation using the Mask R-CNN model. Also, another method using small tiles in semantic segmentation is employed.  Besides that, we are now improving the methods of comparison of the models.

\section{Conclusion}






    In this work, we evaluated models for two different deep learning approaches: semantic segmentation and instance segmentation. For nucleus and NORs segmentation, U-Net with ResNet-18 showed the best results to segment the NORs where the model with ResNet-34 does not show very different results. For nucleus detection and segmentation with instance segmentation, the model with ResNet-50 showed the best results even when compared with a more complex model the ResNet-101.

    Comparing the tested model results from the two approaches, the semantic segmentation with ResNet-18 and ResNet-34 showed similar results and are the best approach for nucleus segmentation considering the IoU metric (0.83). To segment the NORs, the model with ResNet-34 is the best approach with IoU values of 0.92 for clusters and 0.99 for satellite. However, at visual inspection, it was noted that despite a high value in the IoU metric, the models do not segment the satellites very well, but have a good performance for segmenting the NORs (clusters and satellites as a single class). 
    
    Although, the model of instance segmentation shows lower values in the metrics, considering the visual inspection results and the inference velocity this approach can be used as preprocessing. In this way a cascade model can be created, using the instance segmentation to select a nucleus and an semantic segmentation to measure/segment the NORs.
    
    The main contribution of this work is to compare different approaches of automated analysis of cervical cytology slides stained with AgNOR technique using deep learning and show a possible method of automated analysis of this type of images, highlighting the fact that an AgNOR stain is a promising approach for cervical cancer diagnostics and prognostics. We also noticed the need to labeled other classes in the dataset used, to have a wider range of content labeled. We believe that these results can help to spread the AgNOR technique as an alternative to detect cervical cancer at an early stage, improving the treatment success. 


    In future works, we want to test different methods to build a complete pipeline and test the cascade model with instance and semantic segmentation to automated slide analysis to help in the early diagnosis and prognostic of cervical cancer cases. Furthermore, we want to enlarge the size (in image and patients quantities) of the dataset, to help improve the model's assertiveness.

\section*{Acknowledgment}
    We want to thank Coordenação de Aperfeiçoamento de Pessoal de Nível Superior (CAPES) for funding this work.

\bibliographystyle{unsrtnat}
\bibliography{refs}

\end{document}